\documentclass[12pt]{article}
\usepackage{graphicx}
\usepackage{color}

\newcommand{\beq}{\begin{equation}}
\newcommand{\eeq}{\end{equation}}
\newcommand{\bea}{\begin{eqnarray}}
\newcommand{\eea}{\end{eqnarray}}

\newcommand{\gsim}{\lower.7ex\hbox{$
\;\stackrel{\textstyle>}{\sim}\;$}}
\newcommand{\lsim}{\lower.7ex\hbox{$
\;\stackrel{\textstyle<}{\sim}\;$}}

\newcommand{\eod}{\end{document}}

\begin{document}
\thispagestyle{empty}
\vspace*{-22mm}

\begin{flushright}
UND-HEP-16-BIG\hspace*{.08em}02\\
Version 2.0 \\

\end{flushright}
\vspace*{1.3mm}

\begin{center}

{\Large {\bf Bridge between Hadrodynamics \& HEP: Regional CP Violation in Beauty \& Charm Decays}}

\vspace*{5mm}

{ I.I.~Bigi}
\footnote{Will be published in the Proceedings of ICHEP2016}\\
\vspace{5mm}
{\sl Department of Physics, University of Notre Dame du Lac, Notre Dame, IN 46556, USA}\\

{\sl email addresses: ibigi@nd.edu} \\

\vspace*{5mm}

{\bf Abstract}
\vspace*{-1.5mm}
\\
\end{center}
There is a long way from `accuracy' to `precision' about {\bf CP} asymmetries in the decays of beauty \& charm hadrons. 
(a) We have to apply consistent parametrization of the CKM matrix. 
(b) Probing many-body final states (FS) is {\em not} a back-up for understanding the underlying forces; to be realistic we can 
hardly go beyond four-body FS. 
(c) Broken U-spin symmetry is a good tools to describe spectroscopy of hadrons.However the landscape is 
very different for weak transitions; the connection of U- \& V-spin symmetries are important. We have to understand 
the differences between Penguin {\em operators} vs. Penguin {\em diagrams}. 
(d) Collaborations of experimenters \& theorists are crucial with {\em judgment}. 
There is a `hot' news from the conference ICHEP2016: LHCb data show evidence of {\bf CP} asymmetries in the {\bf T}-odd 
moment from $\Lambda_b^0 \to p \pi^-\pi^+\pi^-$. LHCb will follow this `road' 
with $\Lambda_b^0 \to p \pi^-K^+K^-$, $\Lambda_b^0 \to p K^- \pi^+\pi^-$ \& $\Lambda_b^0 \to p K^-K^+K^-$ in run-1. With much more data 
it is crucial to probe its features in {\em regional} asymmetries. \\
A quote from Marinus, who was a $\sim$ 468 AD student of Proklos, a well-known Neoplatonic philosopher: 
"Only {\em being} good is one thing -- but good {\em doing} it is the other one!"

\vspace{3mm}

\hrule

\tableofcontents
\vspace{3mm}

\hrule\vspace{3mm}

\section{{\bf CP} violation beyond the SM}
In the world of "known" matter the SM is at least the leading source of measured {\bf CP} asymmetries in the decays of $K_L$ and 
$B$ mesons. Therefore `we' have to go for `precision' beyond `accuracy'.  
In this short talk I focus mostly about strategies. The central points are: 
(1) We have to use consistent parameterization of the CKM matrix. 
(2) We have to probe {\em many-body} final states (FS). 
(3) The connections between $U$- and $V$-spin (broken) symmetries are very important to understand the 
underlying dynamics, in particular about {\bf CP} asymmetries.  
(4) There is a difference between Penguin {\em operators} and Penguin {\em diagrams}. 
(5) We have to apply more refined tools. Subtle theoretical tools are `waiting'; we have to learn how to apply 
again with  judgment. 
(6) Quark-hadron duality is a subtle tool with its limits. "Duality" is {\em not} an additional assumption; on the other hand often it 
is subtle. I have only the time to mention it  here and there. 
(7) There is a `hot' item: the evidence for {\bf CP} asymmetries in the LHCb data of $\Lambda_b^0 \to p \pi^-\pi^+\pi^-$ 
\&  $\bar \Lambda_b^0 \to \bar p \pi^+\pi^-\pi^+$ \cite{LHCbtalk}. 
To make it shorter: (A) Measuring three- \& four-body FS of charm \& beauty hadrons are {\em not} back-up for 
information from two-body FS -- the landscapes are very different. (B) The best fitted analyses 
often do not give us the best information; i.e., theorists should not be the slaves of the data.

\section{Parameterization of the CKM matrix through ${\cal O}(\lambda^6)$ }
\label{PARA} 
Wolfenstein \cite{WOLFPARA} had put forward a very smart \& successful parametrization with four observables: 
$\lambda \simeq 0.225$ with $A$, $\eta$ \& $\rho$ $\sim {\cal O}(1)$; indeed $A \simeq 0.81$, but $\eta \simeq 0.34$, $\rho \simeq 0.13$ $\ll {\cal O}(1)$. 

Now we need a {\em consistent} parameterization of the CKM matrix with precision as given by \cite{AHN}:
the other three parameters are truly of the order of unity 
($f \sim 0.75$, $\bar h \sim 1.35$ and $\delta_{\rm QM} \sim 90^o$). The SM produces at least the leading source of CPV in measured $B$ transitions:  
\begin{eqnarray}
\nonumber 
V_{\rm CKM}= \left(\footnotesize \begin{array}{ccc} 
V_{ud} & V_{us} & V_{ub} \\
V_{cd} & V_{cs} & V_{cb} \\
V_{td} & V_{ts} & V_{tb}
\end{array}\right)     = 
\; \; \; \; \;  \; \; \; \; \;  \; \; \; \; \; \; \; \; \; \;  \; \; \; \; \;  \; \; \; \; \; \; \; \; \; \; \; \; \; \; \; 
\; \; \; \; \; \; \; \; \; \; \;
&&   \\
=\left(\footnotesize
\begin{array}{ccc}
 1 - \frac{\lambda ^2}{2} - \frac{\lambda ^4}{8} - \frac{\lambda ^6}{16}, & \lambda , & 
 \bar h\lambda ^4 e^{-i\delta_{\rm QM}} , \\
 &&\\
 - \lambda + \frac{\lambda ^5}{2} f^2,  & 
 1 - \frac{\lambda ^2}{2}- \frac{\lambda ^4}{8}(1+ 4f^2) 
 -f \bar h \lambda^5e^{i\delta_{\rm QM}}  &
   f \lambda ^2 +  \bar h\lambda ^3 e^{-i\delta_{\rm QM}}   \\
    & +\frac{\lambda^6}{16}(4f^2 - 4 \bar h^2 -1  ) ,& -  \frac{\lambda ^5}{2} \bar h e^{-i\delta_{\rm QM}}, \\
    &&\\
 f \lambda ^3 ,&  
 -f \lambda ^2 -  \bar h\lambda ^3 e^{i\delta_{\rm QM}}  & 
 1 - \frac{\lambda ^4}{2} f^2 -f \bar h\lambda ^5 e^{-i\delta_{\rm QM}}  \\
 & +  \frac{\lambda ^4}{2} f + \frac{\lambda ^6}{8} f  ,
  &  -  \frac{\lambda ^6}{2}\bar h^2  \\
\end{array}
\right)
+ {\cal O}(\lambda ^7) \; .  
\label{MATRIX}
\end{eqnarray} 
It predicts $\sim$ zero {\bf CP} asymmetries in double Cabibbo decays of charm hadrons 
in the SM  and a {\em maximal} value of 
sin$(2\phi_1) \sim 0.72$ \cite{TROY}. We have to probe correlations with different transitions.

\section{Re-scattering (FSI) \& the Impact of {\bf CPT} invariance}
\label{RESCATT}
The goal is to measure {\bf CP} asymmetries with the impact of New Dynamics (ND), namely their existence  and even their features. They are 
described with amplitudes: 
\begin{equation}
T(P \to f) = e^{i\delta_f} \left[ T_f + 
\sum_{f \neq a_j}T_{a_j}iT^{\rm resc}_{a_jf}  \right] \; \;  ,  \; \; 
T(\bar P \to \bar f) = e^{i\delta_f} \left[ T^*_f +
\sum_{f \neq a_j}T^*_{a_j}iT^{\rm resc}_{a_jf}  \right]   \; ;  
\label{CPTAMP1}
\end{equation}
$T^{\rm resc}_{a_jf}$ describe FSI between $f$ and intermediate 
on-shell states $a_j$ that connect with $f$; $f$ is different from $a_j$, 
but in the same classes of strong dynamics. In the world of quarks 
one describes $a_j = \bar q_j q_j$ and $f= \bar q_k q_k + {\rm pairs \; of \;} \bar q_l q_l$ 
with $q_{j,k,l} = u,d,s$. With{\em out} re-scattering direct {\bf CP} asymmetries cannot happen, even if there are weak phases. 
One gets regional CP asymmetries, not just averaged ones: 
\begin{equation}
\Delta \gamma (f) = |T(\bar P \to \bar f)|^2 - |T(P \to f)|^2 = 
4 \sum_{f \neq a_j} T^{\rm resc}_{a_jf} \, {\rm Im} T^*_f T_{a_j} \; ; 
\label{REGCPV}
\end{equation}
these $f$ consist of two-, three-, four-body etc. states. 
We have to be realistic with finite data and  
a lack of quantitative control of non-perturbative QCD in "acceptable" ways 
\cite{CPBOOK}. 

\subsection{Connections between U- \& V-spin symmetries}
\label{UVSPIN}
 U- \& V-spin symmetries had been introduced to describe spectroscopies of hadrons as subgroups of global $SU(3)_F$  
 before quarks were seen as real physical states. The situation had changed much with {\em weak} transitions. Lipkin suggested based on 
 U-spin symmetry \cite{LIPWEAK}:
\begin{equation}
\Delta_{K\pi} = \frac{A_{CP} (B_d \to K^+\pi^-)}{A_{CP} (B_s \to K^-\pi^+)} + \frac{{\rm BR}(B_s \to K^-\pi^+)}{{\rm BR}(B_d \to K^+\pi^-)}\frac{\tau_d}{\tau_s} = 0 \; ; 
\end{equation}
 2011 data from LHCb gave us \cite{LHCbPRL110}:
\begin{eqnarray}
\nonumber 
A_{\bf CP} (B^0_s \to K^-\pi^+) = 0.27 \pm 0.04 \pm 0.01 &,& A_{\bf CP} (B_d \to K^+\pi^-) = - 0.080 \pm 0.007 \pm 0.03 \\
\Delta_{K\pi}|_{\rm LHCb} &=& - 0.02 \pm 0.05 \pm 0.04
\end{eqnarray}
To get opposite signs for the {\bf CP} violation in the SM is obvious. However, I disagree with this state: 
`These results allow a stringent test of the validity of the  relation between 
$A_{CP} (B_d \to K^+\pi^-)$ \& $A_{CP} (B_s \to K^+\pi^-)$ given':
(1) The value of $\Delta_{K\pi}|_{\rm LHCb}$ very consistent with zero due to U-spin invariance. On the other hand, it is quite consistent also with a value of a few \%, 
as one expects for direct {\bf CP} asymmetry. 
(2) In the world of quarks one `expects' that {\em penguin diagrams} have more impact on $B^0$ than on 
$B_s^0$ transitions.  
(3) One can{\em not} focus only on two-body FS; in particular beauty hadrons produce many-body FS. What about 
{\bf CP} asymmetries in three- \& four-body etc. FS?
(4) The item of quark-duality is actually subtle. 

\subsection{Penguin {\em operators} vs. {\em diagrams} on {\bf CP} violation}
\label{PENGUIN}
The impact of `Penguins' was an important pioneering 1975 work of Shifman, Vainshtein \& Zakharow 
\cite{SVZ75}. It had explained the measured amplitudes of $T(\Delta I =3/2)$ $\ll$ 
$T(\Delta I =1/2)$ in kaon decays; later it was applied to direct {\bf CP} violation in 
Re($\epsilon^{\prime}/\epsilon_K$). It is based on local operators. 

Penguin diagrams can describe suppressed $B$ decays about inclusive {\bf CP} asymmetries with 
hard FSI. However, one cannot do that for exclusive rates 
with soft FSI for hadrons. In special situations we can use other tools like HQE, lattice QCD, 
chiral symmetry, dispersion relations etc. For $\Delta C=1$ transitions one can `draw' Penguin diagrams for SCS decays, but hardly for 
inclusive {\bf CP} violations with local operators and even less for exclusive ones with hadrons. 
`We' have little control over the impact of penguin diagrams in two-body FS for 
$\Delta C \neq 0 \neq \Delta B$. 

\section{{\bf CP} asymmetries in many-body FS}
\label{THREEFOUR}
Probing FS with two hadrons (including narrow resonances) is important to measure {\bf CP} violations; on the other hand one gets `just' numbers. 
However, three- \& four-body FS are described by 
dimensional plots. One needs a lot of work both for experimenters \& theorists, but there might be a prize: to find the existence of ND and even its features. 

\subsection{Dalitz plots of suppressed decays of $B^{\pm}$ mesons}
\label{CHARGEDB}
Data of ${\rm BR}(B^+ \to K^+\pi^+\pi^-) = (5.10 \pm 0.29) \cdot 10^{-5}$ \& 
${\rm BR}(B^+ \to K^+K^+K^-) = (3.37 \pm 0.22) \cdot 10^{-5}$ are not surprising. {\em Averaged} 
{\bf CP} asymmetries \cite{LHCb028}
\begin{eqnarray}
\Delta A_{\bf CP} (B^+ \to K^+\pi^+\pi^-) &=& + 0.032 \pm 0.008 \pm 0.004 \pm 0.007 \\
\Delta A_{\bf CP} (B^+ \to K^+K^+K^-) &=& - 0.043 \pm 0.009 \pm 0.003 \pm 0.007
\end{eqnarray}
are okay for the SM, and it is interesting with opposite signs as {\bf CPT} invariance suggests. 
However look at {\em regional} asymmetries \cite{LHCb028,ERNEST}
\begin{eqnarray}
\Delta A_{\bf CP} (B^+ \to K^+\pi^+\pi^-)|_{\rm regional} &=& 
+ 0.678 \pm 0.078 \pm 0.032 \pm 0.007 \\
\Delta A_{\bf CP} (B^+ \to K^+K^+K^-)|_{\rm regional} &=& 
- 0.226 \pm 0.020 \pm 0.004 \pm 0.007 \; .
\end{eqnarray}
It is very surprising for me due to two connected points: 
The centers of the Dalitz plots are mostly empty and 
the differences are so huge! Can it show the impact of broad resonances like $f_0(500)/\sigma$ and 
$K^*(800)/\kappa$? At least they give us highly non-trivial lessons about non-perturbative QCD.  

Again, no surprises about the rates: 
${\rm BR}(B^+ \to \pi^+\pi^+\pi^-) = (1.52 \pm 0.14) \cdot 10^{-5}$ \& 
${\rm BR}(B^+ \to \pi^+K^+K^-) = (0.50 \pm 0.07) \cdot 10^{-5}$. However look at 
the averaged {\bf CP} asymmetries \cite{IRINA}: 
\begin{eqnarray}
\Delta A_{\bf CP} (B^+ \to \pi^+\pi^+\pi^-) &=& + 0.117 \pm 0.021 \pm 0.009 \pm 0.007 \\
\Delta A_{\bf CP} (B^+ \to \pi^+K^+K^-) &=& - 0.141 \pm 0.040 \pm 0.018 \pm 0.007 \; .
\end{eqnarray}
These number are larger than the other above. Is it surprising that the impact of 
even more suppressed penguin diagrams from the SM is so large? Again looking  
at {\em regional} asymmetries \cite{IRINA,ERNEST}
\begin{eqnarray}
\Delta A_{\bf CP} (B^+ \to \pi^+\pi^+\pi^-)|_{\rm regional} &=& 
+ 0.584 \pm 0.082 \pm 0.027 \pm 0.007 \\
\Delta A_{\bf CP} (B^+ \to \pi^+K^+K^-)|_{\rm regional} &=& 
- 0.648 \pm 0.070 \pm 0.013 \pm 0.007 \; .
\end{eqnarray}
 Having more data is not enough: 
(1) It is crucial {\em not} to stop on two-body FS; measuring three-body FS give us much more 
important information about underlying dynamics.  
(2) {\bf CPT} invariance is still a `usable' tool for analyzing the data. 
(3) The LHCb collaboration defined `good' regional {\bf CP} asymmetries. We 
have to think about that item. Refined tools like dispersion relations will help sizably. 
(4) We have to probe four-body FS.

\subsection{Three- \& four-body FS of charm mesons}
\label{CHARMDECAYS}
{\bf CPT} invariance in charm decays is `practical', since a `few' channels can be combined. 
The SM predicts small averaged asymmetries  for SCS transitions of ${\cal O}(0.1)$\% 
and $\sim$ zero for DCS ones. None has been found yet. 
We have to probe {\em regional}  asymmetries; strong FSI has large impact. 

SCS data give rates for {\em three}-body FS on the scale of several$\times 10^{-3}$ or more that are larger than for two-body FS. In the future we have to probe Dalitz plots 
with the impact of FSI on regional CP asymmetries and their correlations due to {\bf CPT} invariance.
It was discussed in 
Ref.\cite{REIS} with simulations of $D^{\pm} \to \pi^{\pm} \pi^+\pi^-$ and 
$D^{\pm} \to \pi^{\pm} K^+K^-$ with small weak phases and sizable resonances phases in the world of 
hadrons. There are 
good reasons why to compare binned "fractional asymmetries'" vs. "significance" vs. "un-binned" ones \cite{REIS,WILL}. 
For {\em four}-body FS we have rates again on the scale of several$\times 10^{-3}$ or more -- again more than for two-body FS. 

For DCS rates we need huge numbers of charm hadrons; 
PDG15 data set the scales of $10^{-4} - 10^{-3}$ branching ratios.

For four-body FS of charm \& beauty hadrons one can measure the angle $\phi$ between two 
planes of $h_1h_2$ \& $h_3h_4$ and describes to classify its dependence in general \cite{CPBOOK}: 
\begin{eqnarray}
\frac{d\Gamma}{d\phi} (H_Q \to h_1h_2h_3h_4) &=& \Gamma_1 {\rm cos^2}\phi + 
\Gamma_2 {\rm sin^2}\phi +\Gamma_3 {\rm cos}\phi {\rm sin}\phi
\\ 
\frac{d\Gamma}{d\phi} (\bar H_Q \to \bar h_1 \bar h_2 \bar h_3 \bar h_4) &=& \bar \Gamma_1 {\rm cos^2}\phi + \bar \Gamma_2 {\rm sin^2}\phi -\bar \Gamma_3 {\rm cos}\phi {\rm sin}\phi
\end{eqnarray}
The partial widths for 
$H_Q[\bar H_Q] \to h_1h_2h_3h_4[\bar h_1\bar h_2\bar h_3 \bar h_4]$ are given by $\Gamma _{1,2} [\bar \Gamma _{1,2}]$: 
$\Gamma_1 \neq \bar \Gamma_1$ 
and/or  $\Gamma_2 \neq \bar \Gamma_2$ represents direct {\bf CP} violation in the partial widths:
\begin{equation}
\Gamma (H_Q \to h_1h_2h_3h_4) = \frac{\pi}{2} (\Gamma_1 + \Gamma_2)  \; \; \; {\rm vs.} \; \; \; 
\Gamma (\bar H_Q \to \bar h_1\bar h_2 \bar h_3 \bar h_4) = \frac{\pi}{2} (\bar \Gamma_1 + \bar \Gamma_2)  
\end{equation}
$\Gamma_3$ and $\bar \Gamma_3$ represent T {\em odd} correlations \cite{CPBOOK}: 
\begin{equation}
\Gamma_3 \neq \bar \Gamma_3  \; .   
\end{equation}
Integrated rates give $\Gamma_1+\Gamma_2$ vs. $\bar \Gamma_1 + \bar \Gamma_2$; the moments of 
integrated {\em forward-backward} asymmetry 
\begin{equation}
\langle A\rangle = 
\frac{\Gamma_3 - \bar \Gamma_3}{\pi(\Gamma_1+\Gamma_2+\bar \Gamma_1+\bar \Gamma_2)}
\end{equation}
gives information about {\bf CP} violation. When one has enough data to do that, one could disentangle $\Gamma_1$ vs. $\bar \Gamma_1$ and 
$\Gamma_2$ vs. $\bar \Gamma_2$ by tracking the distribution in $\phi$. If there is a {\em production} asymmetry, it gives global 
$\Gamma_1 = c \bar \Gamma_1$, $\Gamma _s = c \bar \Gamma _2$ 
and $\Gamma_3 = - c \bar \Gamma_3$ with {\em global} $c \neq 1$.

\section{{\bf CP} asymmetries in charm \& beauty baryons}
\label{BARYONS}
In principle, {\bf CP} asymmetries have been found in baryons in `our existence'. Back to real world:  
there are huge `hunting regions' for LHCb. Production asymmetries in $pp$ collisions can be calibrated 
by Cabibbo favored decays of $\Lambda_c^+ \to \Lambda \pi^+$ \& $\Lambda_c^+ \to pK^-\pi^+$. 
Thus one can probe {\bf CP} asymmetries in SCS 
$\Lambda_c^+ \to \Lambda K^+$, $p\pi^+\pi^-$, $pK^+K^-$ and in DCS 
$\Lambda_c^+ \to p K^+\pi^-$; furthermore one can -- \& should -- analyze Dalitz plots there. 
So far it was not find in the decays of charm baryons. 

\subsection{`Hot' item: {\bf CP} asymmetry in $\Lambda_b^0$}
\label{HOT}
It would be quite achievement to establish {\bf CP} violation in $\Lambda_b^0$ decays beyond 
production asymmetries in $pp$ collisions. It is very unlikely that these data are connected 
with matter vs. anti-matter asymmetry in our Universe. 
There are several `roads': 
compare $\Lambda_b^0\to p \pi^-$ vs. $\Lambda_b^0\to p K^-$ with 
$\bar \Lambda_b^0\to \bar p \pi^+$ vs. $\bar \Lambda_b^0\to \bar p K^+$ in the rates  
or {\bf T}-odd moments in $\Lambda_b^0 \to p \pi^-\pi^+\pi^-$ \& $\Lambda_b^0 \to p \pi^-K^+K^-$. 
I pointed out at a Belle II workshop B2TTiP at Pittsburgh in May 2016, LHCb meeting at CERN in June and on the first day of ICHEP2016.   
LHCb data \cite{LHCbtalk} give a {\bf T}-odd moment about 
$\Lambda_b^0 \to p \pi^- \pi^+ \pi^-$. This moment is defined by the angle $\phi$ between 
one plane of $\vec p_P$ \& $\vec p_{{\rm fast}\; \pi^-}$ and the other with 
$\vec p_{\pi^+}$ \& $\vec p_{{\rm slow}\; \pi^-}$. It shows a direct {\bf CP} asymmetry with 3.3 $\sigma$ 
uncertainty. It is very interesting. Fig. 4 for Scheme B in Ref.\cite{LHCbtalk} suggests a {\bf CP} asymmetry with $\sim$ 20\% for a {\em regional} asymmetry. 
In run-2 LHCb will probe also $\Lambda_b^0 \to p \pi^-K^+K^-$. 
Much more data will tell us later about the features of the underlying dynamics. 
Furthermore we have to continue with {\bf T}-odd transitions for $\Lambda_b^0 \to p K^- \pi^+\pi^-$ \& $\Lambda_b^0 \to p K^-K^+K^-$.
Can it follow the `road' discussed in Sect.\ref{CHARGEDB}, where penguin diagrams of $b\Rightarrow d$ seems to produce larger impact on 
{\bf CP} asymmetries than $b\Rightarrow s$ ones? Finally we have to think more about the impact of {\em broad} resonances.

\section{Summary of searching for ND in many-body final states}
\label{SUM}
The goal is to find the existence of ND in {\bf CP} asymmetries and maybe also about its features. Now 
there are no `golden' tests of the impact of ND on flavor dynamics. It is crucial to rely on a series 
of arguments with correlations. We need detailed analyses of three- \& four-body FS including {\bf CP} violation, despite the large start-out work. 
The best fitted analyses often do not give us the best information about the underlying dynamics. The tools introduced for analyzing low energy collisions of hadrons by 
hadrodynamics (like dispersion relations) are crucial to go from accuracy to precision and find ND as non-leading source.

\vspace{2mm}

{\bf Acknowledgments:} This work was supported by the NSF under PHY-1520966.

\end{document}